\documentclass[3p,times]{elsarticle}

\usepackage{bm}
\usepackage{color}
\usepackage{ulem}

\usepackage{amssymb}

\journal{Computer Physics Communications}

\begin{document}

\begin{frontmatter}



\title{
Polynomial Expansion Monte Carlo Study of Frustrated Itinerant Electron Systems: Application to a Spin-ice type Kondo Lattice Model on a Pyrochlore Lattice
}


\author[utokyo]{Hiroaki Ishizuka}
\author[utokyo,mpipks]{Masafumi Udagawa}
\author[utokyo]{Yukitoshi Motome}

\address[utokyo]{
Department of Applied Physics, University of Tokyo, Tokyo, 113-8656, Japan
}
\address[mpipks]{
Max Planck Institute for the Physics of Complex Systems, D-01187 Dresden, Germany
}

\begin{abstract}
We present the benchmark of the polynomial expansion Monte Carlo method to a Kondo lattice model
with classical localized spins on a geometrically frustrated lattice. 
The method enables to reduce the calculation amount by using the Chebyshev polynomial expansion 
of the density of states compared to a conventional Monte Carlo technique 
based on the exact diagonalization of the fermion Hamiltonian matrix. 
Further reduction is brought by a real-space truncation of the vector-matrix operations. 
We apply the method to the model with spin-ice type Ising spins on a three-dimensional pyrochlore lattice, 
and carefully examine the convergence in terms of the order of polynomials and the truncation distance.
We find that, in a wide range of electron density at a relatively weak Kondo coupling compared to the
noninteracting bandwidth, the results by the polynomial expansion method show good convergence to those
by the conventional method within reasonable numbers of polynomials.
This enables us to study the systems up to $4 \times 8^3 = 2048$ sites, while the previous study by the
conventional method was limited to $4 \times 4^3 = 256$ sites.
On the other hand, the real-space truncation is not helpful in reducing the calculation amount for the system
sizes that we reached, as the sufficient convergence is obtained when most of the sites are involved within
the truncation distance.
The necessary truncation distance, however, appears not to show significant system size dependence, suggesting
that the truncation method becomes efficient for larger system sizes.
\end{abstract}

\begin{keyword}
Monte Carlo method \sep polynomial expansion method \sep geometrical frustration \sep Kondo lattice model \sep spin ice \sep pyrochlore lattice
\end{keyword}

\end{frontmatter}


\section{Introduction}

Interplay between localized spins and itinerant electrons has been one of the major topics in the field of strongly correlated electrons.
Localized spins considerably affects the charge degree of freedom of itinerant electrons, leading to fascinating transport phenomena, such as non Fermi liquid behavior in the quantum critical region in rare-earth compounds~\citep{Stewart2001} and the colossal magneto-resistance (CMR) in perovskite manganese oxides~\citep{Dagotto2002}.
On the other hand, for the localized spin degree of freedom, the kinetic motion of itinerant electrons results in effective interactions between localized spins, such as the Ruderman-Kittel-Kasuya-Yosida (RKKY) interaction~\citep{RKKYInteraction} and double-exchange (DE) interaction~\citep{DEMechanism}.
These effective interactions give rise to peculiar magnetic orderings in the spin-charge coupled systems. 
Hence, in these systems, the charge and spin degrees of freedom affect each other in an entangled way, and it is not allowed to treat only one degree of freedom by fixing the other.
The equilibrium state is obtained by optimizing the total free energy of the system. 

The study on the spin-charge coupling has recently been extended to frustrated magnetic conductors in which itinerant electrons are coupled to localized spins on geometrically frustrated lattices.
Experimentally, metallic pyrochlore oxides have attracted considerable attention~\citep{Gardner2010}, in which itinerant $d$ electrons interact with localized $f$ electrons. 
Theoretically, however, geometrical frustration brings further difficulty into the problem of spin-charge coupled systems. 
First, in the presence of frustration, the coupling between itinerant electrons and localized quantum spins leads to the negative sign problem, which hampers precise calculations at low temperatures. 
Meanwhile, when the localized spins are approximated as classical vectors and the electrons do not have any direct interaction between them (the situation that we focus on in the present work), the negative sign problem can be avoided.
Even in this simplified situation, however, systematic calculations are still challenging; in order to examine the effect of large fluctuations, it is desired to make numerical simulations, but the calculation amount is usually large and increases rapidly as increasing system sizes.

For the spin-charge coupled systems with classical localized spins, unbiased Monte Carlo (MC) calculations have been developed and used in the field of CMR manganites in which the geometrical frustration does not play a major role~\citep{Furukawa1999}. 
Originally, the MC method was implemented by using the exact diagonalization (ED) for the electronic degree of freedom to calculate the MC weight for sampling the configuration of the classical spins~\citep{Yunoki1998}.
We call this algorithm EDMC. 
Later, several alternative methods were proposed for reducing the calculation amount~\citep{Motome1999,Furukawa2001,Furukawa2004,Alvarez2005,Alonso2001,Weisse2009}. 
One of the sophisticated methods is the polynomial expansion MC (PEMC) method proposed by one of the authors and his collaborators~\citep{Motome1999,Furukawa2001}.
In this method, the polynomial expansion of the density of states (DOS) is used to calculate the MC weight, which reduces the calculation amount than ED; the method was further improved by introducing the truncation in the vector-matrix operations~\citep{Furukawa2004,Alvarez2005}.
Hence, PEMC allows analyses on much larger system sizes compared to EDMC~\citep{Motome2000,Motome2001a,Motome2001b,Motome2003a,Motome2003,Sen2006}. 

Because of the growing interests in frustrated magnetic conductors, it is desired to apply such MC simulations to the models with geometrical frustration. 
The application of PEMC, however, is anticipated to be less efficient for frustrated systems as, in general, DOS has a singular form under frustration.
For example, a $\delta$-function peak can appear associated with a flat band in the case of a kagome (corner-sharing triangles) and a pyrochlore (corner-sharing tetrahedra) lattice.
A large number of polynomial bases are required to reproduce such a $\delta$ function in DOS with high accuracy, implying slow convergence in the PEMC framework and larger computational amount.
Another difficulty is that the spin-charge coupling in frustrated magnetic conductors is often very weak compared to the large Hund's-rule coupling in CMR materials, and hence, the RKKY mechanism is dominant to make the effective spin interactions long-ranged and oscillating. 
This makes the truncation scheme less efficient since subtle energy differences originating from further-neighbor interactions play a decisive role.

Due to these anticipated problems, MC studies of frustrated spin-charge coupled systems have been very limited so far. 
For instance, the DE model on a pyrochlore lattice was studied, but the accessible system sizes were small since the studies were made by using EDMC~\citep{Motome2010-1,Motome2010-2,Motome2010-3}.
Models in two dimensions were also studied on a triangular lattice~\citep{Kumar2010,Kato2010,Ishizuka2012-2} and kagome lattice~\citep{Ishizuka2012-3}, using EDMC.
While PEMC was recently applied to a triangular lattice model, the studies focused on the case with a large Hund's-rule coupling which is comparable to the bandwidth~\citep{Zhang2011,Zhang2012}.
In addition, DOS of the triangular lattice model in the noninteracting limit does not have a flat band, in contrast to the kagome and pyrochlore models.
Hence, it does not involve the difficulties due to the $\delta$-function structure in DOS.

Recently, the authors applied PEMC to a model for metallic pyrochlore oxides, i.e., a spin-ice type Kondo lattice model on a pyrochlore lattice~\citep{Ishizuka2012}. 
They calculated the magnetic and electronic properties of the three-dimensional model in a relatively weak-coupling regime up to $4 \times 8^3$ sites systematically, and clarified the phase diagram including a variety of different phases.
In the course of the study, the applicability of PEMC was examined carefully, and a part of the benchmark was reported in Ref.~\citep{Ishizuka2012-4}.
The details, however, have not been discussed yet.

In this paper, we present the detailed benchmark on the application of PEMC to the spin-ice type Kondo lattice model on a pyrochlore lattice.
We focus on the low electron density region in the relatively weak-coupling regime, which is considered to be relevant for the metallic pyrochlore oxides.
We perform the benchmark on the convergence of PEMC in terms of the order of polynomials and the truncation of vector-matrix operations by calculating both long-range order parameters and local correlations. 
The results indicate that although DOS has a singular form including a $\delta$-function peak in the noninteracting limit, PEMC turns out to be applicable in a wide range of electron density except for the very low density region.
The required order of polynomials is within the range comparable to the typical numbers in the previous studies for unfrustrated models.
On the other hand, although the weak spin-charge coupling makes the truncation less helpful to reduce the calculation amount for currently accessible systems sizes, our benchmark suggests that it will become efficient in the simulations on larger size systems than those used in the present study. 

The organization of this paper is as follows.
In Sec.~\ref{sec:model}, we introduce the model and method. 
The model Hamiltonian and parameters are described in Sec.~\ref{sec:model_and_parameters}.
A brief introduction on the PEMC method is given in Sec.~\ref{subsec:pem} and Sec.~\ref{subsec:truncation}.
In Sec.~\ref{sec:simulation}, we present the results of the benchmark. 
In Sec.~\ref{sec:sim:Tdep}, we show the temperature dependence of the sublattice magnetization calculated by EDMC and PEMC.
Detailed comparison of PEMC to EDMC for $4\times 4^3$ site systems is shown in Sec.~\ref{sec:sim:polynomial} and Sec.~\ref{sec:sim:truncation}.
The results for larger system sizes are presented in Sec.~\ref{sec:sim:largesize}.
Discussions on the results are elaborated in Sec.~\ref{sec:discussion}.
Finally, Sec.~\ref{sec:summary} is devoted to the summary.

\section{Model and method} \label{sec:model}

In this section, we introduce the model and method used in our benchmark study in the following sections. 
We use the PEMC method developed in the previous studies~\citep{Motome1999,Furukawa2004,Alvarez2005} with a minor modification in the truncation method. 
The procedure of PEMC is briefly reviewed to make the paper self-contained.

\subsection{Model and parameters} \label{sec:model_and_parameters}

We consider a Kondo lattice model with Ising spins on a pyrochlore lattice~\citep{Ishizuka2012,Udagawa2012}, whose Hamiltonian is given by
\begin{eqnarray}
H = -t \! \sum_{\langle i,j \rangle, \sigma} \! ( c^\dagger_{i\sigma} c_{j\sigma} + {\rm H.c.} ) -J \sum_{i, \sigma, \sigma^\prime} c_{i\sigma}^\dagger {\boldsymbol \sigma}_{\sigma \sigma^\prime} 
c_{i\sigma^\prime} \cdot {\bf S}_i.
\label{eq:H}
\end{eqnarray}
The first term represents hopping of itinerant electrons, where $c_{i\sigma}$ ($c^\dagger_{i\sigma}$) is the annihilation (creation) operator of an itinerant electron with spin $\sigma= \uparrow, \downarrow$ at $i$th site, and $t$ is the transfer integral.
The sum $\langle i,j \rangle$ is taken over nearest-neighbor sites on the pyrochlore lattice which consists of a three-dimensional network of corner-sharing tetrahedra [see Figs.~\ref{fig:sim:order}(d)-(f)].
The second term is the onsite interaction between itinerant electron spins (${\bf \sigma}$ is the Pauli matrix) and localized Ising spins ${\bf S}_i$ ($|{\bf S}_i|=1$), and $J$ is the coupling constant (the sign of
$J$ does not matter, since the localized spins are classical). 
The anisotropy axis of Ising spin is given along the local $\langle 111\rangle$ direction, i.e., along the line connecting the centers of two tetrahedra which the spin belongs to [see Figs.~\ref{fig:sim:order}(d)-(f)].

Considering the situations in many pyrochlore oxides, we focus on the relatively low electron density region of $0 < n < 0.35$ at a weak spin-charge coupling $J=2t$ compared to the noninteracting bandwidth $8t$.
Here, the electron density is defined by 
\begin{eqnarray}
n=\frac1{2N} \sum_{i \sigma}\langle c_{i\sigma}^\dagger c_{i \sigma}\rangle,
\end{eqnarray}
where $N$ is the number of sites.
Hereafter, we set the unit of energy $t=1$, the lattice constant of the cubic unit cell $a = 1$, and the Boltzmann constant $k_{\rm B} = 1$.

\subsection{Polynomial expansion Monte Carlo method} \label{subsec:pem}

In the model in eq.~(\ref{eq:H}), itinerant electrons have no direct interaction between them, but coupled only to the classical Ising spins.
The model belongs to the category of models in which noninteracting fermions couple with classical fields. 
In general, the partition function for such models is obtained by taking two traces; one is over the classical fields and the other over the fermion degree of freedom.
For the present model (\ref{eq:H}), the partition function is written as
\begin{eqnarray}
Z = {\rm Tr}_{\{ {\bf S}_i \}} {\rm Tr}_{\{ c_{i\sigma}, c_{i\sigma}^\dagger \}}
\exp\left[-\beta \left( H(\{ {\bf S}_i \}) - \mu \hat{N}_c \right)\right], 
\end{eqnarray}
where ${\rm Tr}_{\{ {\bf S}_i \}}$ and ${\rm Tr}_{\{ c_{i\sigma}, c_{i\sigma}^\dagger \}}$ are the traces over
the Ising spins and the electron operators, respectively, and $H(\{ {\bf S}_i \})$ is a one-particle Hamiltonian matrix in eq.~(\ref{eq:H}) defined for a given Ising spin configuration $\{ {\bf S}_i \}=({\bf S}_1, {\bf S}_2, \cdots, {\bf S}_N)$; $\beta=1/T$ is inverse temperature, $\mu$ is the chemical potential, and $\hat{N}_c =\sum_{i\sigma} c_{i \sigma}^\dagger c_{i \sigma}$.
The former trace can be calculated by classical MC sampling of the spin configurations $\{{\bf S}_i\}$ with the Boltzmann weight
\begin{eqnarray}
P(\{{\bf S}_i\})=\frac{1}{Z} \exp \left[-S_{\rm eff}(\{{\bf S}_i\})\right],
\end{eqnarray}
where the effective action is given by the latter trace in the form
\begin{eqnarray}
S_{\rm eff}(\{{\bf S}_i\})=-\log\left({\rm Tr}_{\{ c_{i\sigma}, c_{i\sigma}^\dagger \}} \exp\left[-\beta \left( H(\{ {\bf S}_i \}) - \mu \hat{N}_c \right) \right] \right). 
\end{eqnarray}

A straightforward method to calculate the effective action is ED of $H(\{ {\bf S}_i \})$, which is used in EDMC~\citep{Yunoki1998}.
By using the one particle eigenvalues for $H(\{ {\bf S}_i \})$, $\{\varepsilon_\nu(\{{\bf S}_i\})\}$, the effective action is calculated by
\begin{eqnarray}
S_{\rm eff}(\{{\bf S}_i\}) = \sum^{N_{\rm dim}}_{\nu=1} F\left(\varepsilon_\nu(\{{\bf S}_i\}) \right), 
\end{eqnarray}
where $F(x) = - \log \left[1+\exp\left\{-\beta (x - \mu)\right\}\right]$ and $N_{\rm dim}$ is the
dimension of the Hamiltonian ($N_{\rm dim} = 2N$ in the present case).

In the PEMC method, the sum over the eigenstates is replaced by the integration over DOS,
and the integral is evaluated by using the polynomial expansion technique~\citep{Motome1999}; 
\begin{eqnarray}
S_{\rm eff}(\{{\bf S}_i\}) = \int d\varepsilon \, D_{\{{\bf S}_i\}}(\varepsilon) F(\varepsilon)
                           = \sum_m \mu_m f_m,\label{eq:Seff7}
\end{eqnarray}
where $D_{\{{\bf S}_i\}}$ is DOS for itinerant electrons for a spin configuration $\{{\bf S}_i\}$.
In eq.~(\ref{eq:Seff7}), DOS and $F$ are expanded by Chebyshev polynomials as
\begin{eqnarray}
&& \mu_m = \int_{-1}^1 dx T_m(x) \tilde{D}_{\{{\bf S}_i\}}(x)
= {\rm Tr}\, T_m\left( H(\{ {\bf S}_i \}) \right), \label{eq:mu_m} \\
&& f_m = \frac{-1}{\alpha_m} \int_{-1}^1 \frac{dx}{\pi \sqrt{1-x^2}} T_m(x) F(x),
\label{eq:f_m}
\end{eqnarray}
where $\alpha_m =1$ for $m=0$ and otherwise $1/2$. 
Here, DOS is renormalized so that the entire spectrum fits into the range of $x=[-1,1]$;
\begin{eqnarray}
\tilde{D}_{\{{\bf S}_i\}}(x) =  aD_{\{{\bf S}_i\}}(ax+b),
\end{eqnarray}
where $a= (\varepsilon_{\rm top} - \varepsilon_{\rm btm})/2$ and $b= (\varepsilon_{\rm top} + \varepsilon_ {\rm btm})/2$ with $\varepsilon_{\rm top} = 2t+J+1$ and $\varepsilon_{\rm btm} = -6t-J-1$ (we afford a margin of $1$ for both $\varepsilon_{\rm top}$ and $\varepsilon_{\rm btm}$).
In eqs.~(\ref{eq:mu_m}) and (\ref{eq:f_m}), $T_m$ is the Chebyshev polynomials defined in a recursive form as
$T_0(x)=1$, $T_1(x)=x$, 
and
$T_m(x)= 2xT_{m-1}(x) - T_{m-2}(x)$. 

In the MC update in PEMC, the Chebyshev moment $\mu_m$ is evaluated by calculating the Chebyshev
polynomials of the Hamiltonian matrix recursively.
For the sparse Hamiltonian matrix, the calculation amount of $\mu_m$ is $O(N^2 \log N)$, as the necessary order of polynomials scales as $\log N$.
Hence, the total cost for one MC update in PEMC is $O(N^3 \log N)$~\citep{Motome1999}, which is reduced from $O(N^4)$ in EDMC.

\subsection{Truncation algorithm}  \label{subsec:truncation}

To further reduce the calculation amount, one of the authors and his collaborator proposed a truncation algorithm~\citep{Furukawa2004}. 
In the truncation procedure, a real-space basis ${\bf e}_{j}(k) = \delta_{j,k}$ is chosen for the trace in eq.~(\ref{eq:mu_m}), where $k$ is a site index.
A new vector ${\bf v}_{j}^{(m)}$ is generated by multiplying the unit vector by the $m$th Chebyshev polynomial of the Hamiltonian, as
\begin{eqnarray}
{\bf v}_{j}^{(m)} =T_m\left( H(\{ {\bf S}_i \}) \right) {\bf e}_{j} \equiv\sum_{k} v_{j,k}^{(m)} {\bf e}_{k}.
\label{eq:vector} 
\end{eqnarray}
If the hopping term in the Hamiltonian is limited to nearest-neighbor sites as in eq.~(\ref{eq:H}), the coefficient $v_{j,k}^{(m)}$ takes a nonzero value only if $||j-k|| \le m$ is satisfied, where $||j-k||$ is the Manhattan distance between two sites $j$ and $k$.
Furthermore, the coefficient usually becomes small quickly as the Mahnattan distance increases.
Hence, the vector elements of ${\bf v}_j^{(m)}$ with such small amplitudes can be neglected in the calculation of the moment $\mu_m$. 
In particular, the truncation was done by introducing a threshold for the amplitude of vector elements, $\epsilon$, and ignoring the small elements which satisfy $| v_{j,k}^{(m)} |< \epsilon$ in the calculation of eq.~(\ref{eq:vector})~\citep{Furukawa2004}. 
A similar truncation was also introduced in the trace operation to calculate the effective action $S_{\rm eff}(\{{\bf S}_i\})$.
This algorithm further reduces the total cost of one MC update to $O(N)$~\citep{Furukawa2004}.

\begin{figure}[tbhp]
\begin{center}
\includegraphics[width=0.36\linewidth]{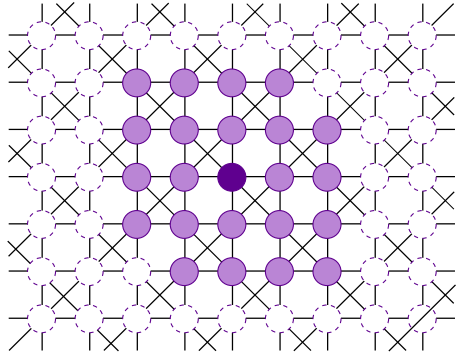}
\end{center}
\caption{
(color online). 
Schematic picture of the real-space truncation.
The figure shows a projection of the pyrochlore lattice onto a $\langle 001 \rangle$ plane, 
and circles represent the lattice sites. 
The dark circle in the center represents the site with flipped spin.
Light-colored circles indicate the sites within the range of the truncation distance $d$. 
The picture shows an example of $d=2$.
}
\label{sim:fig:tpem}
\end{figure}

In this paper, we test the efficiency of a similar but slightly different truncation algorithm for the model in eq.~(\ref{eq:H}).
We here carry out the truncation by a real-space distance, not by a magnitude of the vector element in the original scheme; namely, we set a truncation distance $d$ and ignore all contributions out of the range of  the Manhattan distance $d$ from a flipped spin (Fig.~\ref{sim:fig:tpem}).
In the present method, the list of sites to be considered in the calculation is known in advance and unchanged throughout the MC simulation.
On the other hand, in the previous method, the list needs to be updated by looking at the elements of ${\bf v}_{j}^{(m)}$ in each MC step.
Therefore, the present algorithm is much simpler than the previous one.

\subsection{Monte Carlo details and physical quantities}

PEMC is a well-controlled approximation in the sense that the results converge to the EDMC results when one takes the polynomial expansion up to a sufficiently high order and sufficiently long truncation length.
In the following sections, we present the benchmark results for such convergence by changing the total number of polynomials $m_{\rm tot}$ and the truncation distance $d$.
The calculations were conducted up to $N=4\times 8^3$ for various temperatures and electron fillings.
We also performed EDMC for $N=4\times 4^3$ to compare with the PEMC results.
Typically, we performed 3000 MC measurements after 500 MC steps of thermalization. 
One MC update for the system size $N=4\times 8^3$ with $m_{\rm tot}=40$ and no truncation takes about 50 seconds by using 1024 CPU cores in the System B (SGI Altix ICE 8400EX) at ISSP supercomputer center. 

In the benchmark study, we measure two types of physical quantities which characterize the long-range ordering and short-range correlations, respectively.
For the long-range ordering, we calculate the sublattice magnetization
\begin{eqnarray}
M_{\bf q} =  \left[ \frac{4}{N} S^{\alpha\alpha}({\bf q}) \right]^{1/2},
\end{eqnarray}
where $S^{\alpha\alpha}({\bf q})$ is the $\alpha$th diagonal component of the spin structure factor.
Here, the spin structure factor is given by
\begin{eqnarray}
S^{\alpha\beta}({\bf q}) = \frac{1}{N} \sum_{n,l} \langle {\bf S}_n^\alpha \cdot {\bf S}_l^\beta \rangle \exp [ {\rm i}{\bf q} \cdot ({\bf r}_n^\alpha - {\bf r}_l^\beta)],
\end{eqnarray}
where ${\bf S}_n^\alpha$ is the classical Ising spin at $\alpha$th sublattice site in $n$th unit cell, and
${\bf r}_n^\alpha$ is the position vector of the Ising spin ${\bf S}_n^\alpha$. 
Here, $\bf q$ represents the characteristic wave number for the low-temperature magnetic structure, which depends on the electron density $n$ as well as $J$. 
In the calculations below, we present the results for three magnetic phases: ice-ferro, 32-sublattice, and all-in/all-out order~\citep{Ishizuka2012} (see Sec.~\ref{sec:simulation} for the details).
The characteristic wave number $\bf q$ is given by ${\bf q}=(0,0,0)$ for the ice-ferro and all-in/all-out orders, and ${\bf q}=(\pi,\pi,\pi)$ for the 32-sublattice order.
As the diagonal component of the structure factor $S^{\alpha\alpha} ({\bf q})$ does not depend on $\alpha$ for these three orders, we show the results for the sublattice $\alpha=1$.

In addition to the sublattice magnetization, we also measure short-range correlations. 
Here we use the local correlation parameters $P_{22}$, $P_{31}$, and $P_{40}$ defined by the probabilities of a tetrahedron to be in two-in two-out configuration, three-in one-out or one-in three-out configuration, and all-in or all-out configuration, respectively.
The magnetic ordering pattern is determined by these local correlation parameters in addition to the spin structure factor.

\section{Benchmark result} \label{sec:simulation}

In this section, we present the benchmark of PEMC on the spin-ice type Kondo lattice model in eq.~(\ref{eq:H}) at $J=2t$.
In Sec.~\ref{sec:sim:Tdep}, we present overall $T$ dependences of the sublattice magnetization, $M_{\bf q}$, while varying the total number of polynomials $m_{\rm tot}$ for $N=4\times 4^3$ at several electron fillings with $d=6$.
The detailed analysis of $M_{\bf q}$ and the local correlation parameters is discussed for the convergence with respect to $m_{\rm tot}$ and the truncation distance $d$ in Sec.~\ref{sec:sim:polynomial} and Sec.~\ref{sec:sim:truncation}, respectively.
Results for larger system sizes are presented in Sec.~\ref{sec:sim:largesize}.

\subsection{Temperature dependence of sublattice magnetization} \label{sec:sim:Tdep}

\begin{figure}[tbhp]
\begin{center}
\includegraphics[width=0.54\linewidth]{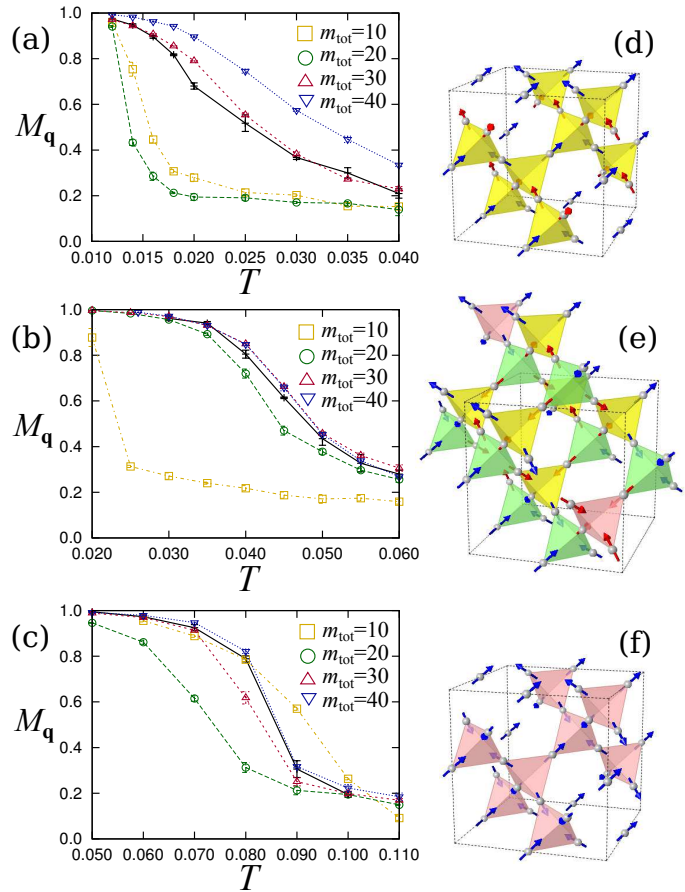}
\end{center}
\caption{
(color online). 
$T$ dependences of $M_{\bf q}$ calculated by PEMC at (a) $\mu = -5.9$ [$n=0.030(2)$], (b) $\mu = -3.7$
[$n=0.180(5)$], and (c) $\mu = -1.3$ [$n=0.348(6)$].
The wave number ${\bf q}$ is ${\bf q}=(0,0,0)$ for (a) and (c), and ${\bf q} = (\pi,\pi,\pi)$ for (b).
Different symbols correspond to the results of MC simulation with different numbers of polynomials.
Crosses with solid lines show the results by EDMC.
The PEMC calculations are done with $d=6$ for $4\times 4^3$ site systems.
The right figures depict the magnetic structures for (d) ice-ferro, (e) 32-sublattice, and (f) all-in/all-out orders, which correspond to the low-$T$ states in the data in (a), (b), and (c), respectively.
}
\label{fig:sim:order}
\end{figure}

Recently, the authors have conducted a systematic study on the phase diagram of the spin-ice Kondo lattice model in eq.~(\ref{eq:H})~\citep{Ishizuka2012}. 
The results indicate that four dominant magnetic phases arise in the electron density $n\lesssim 0.3$:
ice-ferro [Fig.~\ref{fig:sim:order}(d)], ice-$(0,0,2\pi)$, 32-sublattice [Fig.~\ref{fig:sim:order}(e)], and all-in/all-out [Fig.~\ref{fig:sim:order}(f)].
To gain an overview on how PEMC works, we first show $T$ dependences of the sublattice magnetization $M_{\bf q}$ in the ice-ferro, 32-sublattice, and all-in/all-out ordered regions.
The data were calculated for $N=4\times 4^3$ with $d=6$.
As we show later, the deviation due to the truncation is sufficiently small for $d=6$.
We omit the results for the ice-$(0,0,2\pi)$ phase as PEMC shows extremely slow convergence in terms of the number of polynomials $m_{\rm tot}$.

Figure~\ref{fig:sim:order}(a) shows $T$ dependence of $M_{\bf q}$ calculated by PEMC at $\mu = -5.9$ in comparison with the EDMC result. 
The electron density is almost $T$ independent at $n=0.030(2)$~\citep{note_on_density}.
In this very low density region, the ice-ferro order develops at low $T$, whose ordering pattern is shown in Fig.~\ref{fig:sim:order}(d). 
The PEMC results are shown by the open symbols, while the EDMC results are shown by crosses connected by the solid line.
All the results show a rapid increase of $M_{\bf q}$ as $T$ decreases, signaling the phase transition to the ice-ferro phase.
However, the PEMC results show slow convergence to the EDMC ones; even the results for $m_{\rm tot}=40$ show considerable deviations.

On the other hand, for higher electron densities, the PEMC results show good convergence to the EDMC results.
Figures~\ref{fig:sim:order}(b) and \ref{fig:sim:order}(c) show the results at $\mu=-3.7$ and $\mu=-1.3$, respectively.
In the intermediate density region in Fig.~\ref{fig:sim:order}(b), the 32-sublattice order is stabilized at low-$T$ [see Fig.~\ref{fig:sim:order}(e)], while for the higher density in Fig.~\ref{fig:sim:order}(c), the all-in/all-out order appears [see Fig.~\ref{fig:sim:order}(f)].
As shown in Fig.~\ref{fig:sim:order}(b), at $\mu=-3.7$, while the results for $m_{\rm tot}=10$ and $20$ show considerable deviations from the EDMC results, the results for $m_{\rm tot}=30$ and $40$ show good agreement except for a slight deviation in the critical region (the critical temperature is estimated as $T_c = 0.043(2)$ from the inflection point of the $T$ dependence of $M_{\bf q}$~\citep{Ishizuka2012}).
The situation is similar for $\mu=-1.3$ in Fig.~\ref{fig:sim:order}(c); $m_{\rm tot}=30$ and $40$ appear to be enough for the convergence except for the critical region near $T_c=0.085(5)$, whereas the results for smaller $m_{\rm tot}$ are oscillating in terms of $m_{\rm tot}$.

These results indicate that the PEMC results show sufficient convergence for $m_{\rm tot} \gtrsim 30$ in the relatively high density region of $n \gtrsim 0.15$. 
The results away from the critical region converge faster than those in the critical region.
These aspects are further discussed in the next section.  

\subsection{Convergence in terms of the number of polynomials} \label{sec:sim:polynomial}

\begin{figure}[tbhp]
\begin{center}
\includegraphics[width=0.6\linewidth]{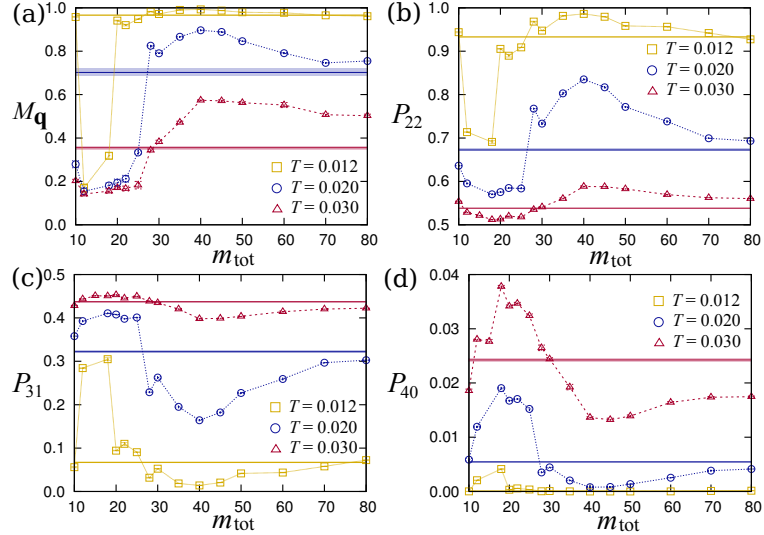}
\end{center}
\caption{
(color online). 
PEMC results for $m_{\rm tot}$ dependences of (a) $M_{\bf q}$, (b) $P_{22}$, (c) $P_{31}$, and (d) $P_{40}$. 
The calculations are done for $N=4\times 4^3$ and $\mu=-5.7$ with $d=6$.
For comparison, the results and statistical errors by EDMC are shown by horizontal solid lines and shades, respectively.
}
\label{fig:sim:npoly57m}
\end{figure}

We investigate the convergence of PEMC with respect to $m_{\rm tot}$ in the three density regions shown in Figs.~\ref{fig:sim:order}(a)-(c), respectively, for $N=4\times 4^3$.
We here show the convergence in three different $T$ regions: a low-$T$ ordered region, high-$T$ paramagnetic region, and in the vicinity of the critical temperature.
The critical temperatures for $\mu=-5.7$, $-3.7$, and $-1.3$ are estimated to be $T_c=0.023(2)$, $0.043(2)$, and $0.085(5)$, respectively, from the inflection point of the $T$ dependence of $M_{\bf q}$~\citep{Ishizuka2012}.

Figure~\ref{fig:sim:npoly57m} shows the results for $\mu = -5.7$ (ice-ferro ordered region) at $T=0.012$, $0.020$, and $0.030$, corresponding to the ordered, critical, and paramagnetic regions, respectively.
Figure~\ref{fig:sim:npoly57m}(a) is the result for $M_{\bf q}$, and Figs.~\ref{fig:sim:npoly57m}(b)-\ref{fig:sim:npoly57m}(d) are the results for the local correlation parameters $P_{22}$, $P_{31}$, and $P_{40}$, respectively.
The EDMC results are indicated by horizontal solid lines (the error bars are shown by shades).
In all $T$ regions, the convergence of PEMC results to the EDMC ones is slow; $m_{\rm tot} \gtrsim 80$ appears to be necessary for sufficient convergence in this low density region.

\begin{figure}[tbhp]
\begin{center}
\includegraphics[width=0.6\linewidth]{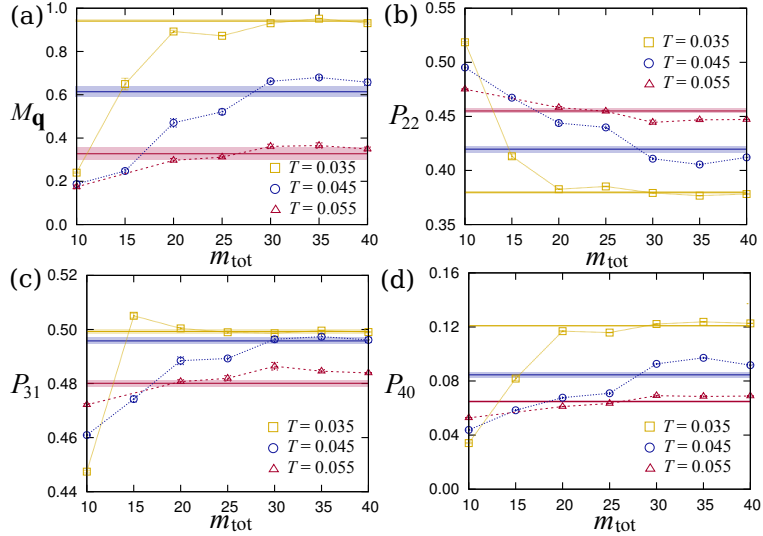}
\end{center}
\caption{
(color online). 
PEMC results for $m_{\rm tot}$ dependences of (a) $M_{\bf q}$, (b) $P_{22}$, (c) $P_{31}$, and (d) $P_{40}$. 
The calculations are done for $N=4\times 4^3$ and $\mu=-3.7$ with $d=6$.
For comparison, the results and statistical errors by EDMC are shown by horizontal solid lines and shades, respectively.
}
\label{fig:sim:npoly37m}
\end{figure}

\begin{figure}[tbhp]
\begin{center}
\includegraphics[width=0.6\linewidth]{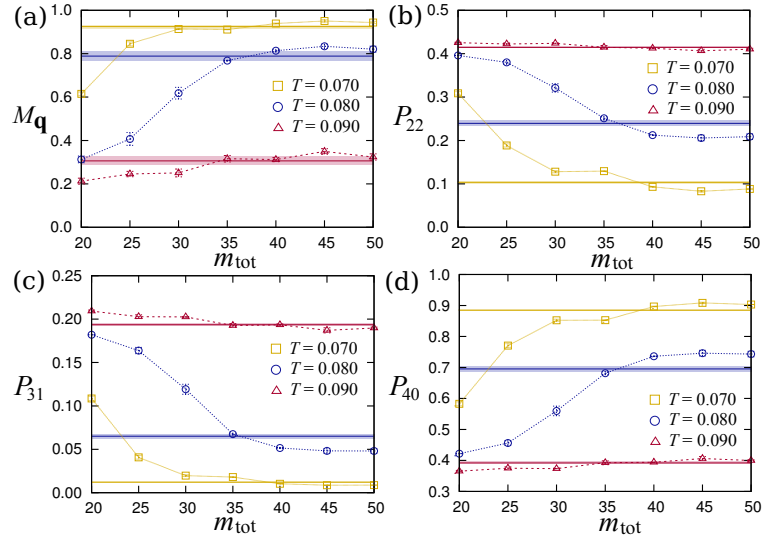}
\end{center}
\caption{
(color online). 
PEMC results for $m_{\rm tot}$ dependences of (a) $M_{\bf q}$, (b) $P_{22}$, (c) $P_{31}$, and (d) $P_{40}$. 
The calculations are done for $N=4\times 4^3$ and $\mu=-1.3$ with $d=6$.
For comparison, the results and statistical errors by EDMC are shown by horizontal solid lines and shades, respectively.
}
\label{fig:sim:npoly13m}
\end{figure}

On the other hand, the PEMC results at $\mu=-3.7$ (the 32-sublattice ordered region) show better convergence.
The results are shown in Fig.~\ref{fig:sim:npoly37m}.
Both the sublattice magnetization  and the local correlation parameters show reasonable convergence for $m_{\rm tot} \gtrsim 30$.
The situation is similar for $\mu=-1.3$ (the all-in/all-out ordered region), as shown in Fig.~\ref{fig:sim:npoly13m}.
The results show good convergence  for $m_{\rm tot} \gtrsim 35$.
In both cases with $\mu=-3.7$ and $-1.3$, the data in the critical region appear to show relatively slower convergence compared to the low-$T$ and high-$T$ region, but the remnant deviation is in a reasonable range and not harmful to the estimation of the critical temperature~\citep{Ishizuka2012}.

The results in Figs.~\ref{fig:sim:npoly37m} and \ref{fig:sim:npoly13m} indicate that PEMC works efficiently in a wide range of $T$ for the relatively high electron density of $n \gtrsim 0.15$.
Typically, $m_{\rm tot}=30$-$40$ is enough for the convergence.
On the other hand, in the lower electron density region, quantitatively sufficient convergence requires much larger $m_{\rm tot}$.
These points are discussed in Sec.~\ref{sec:discussion}.

\subsection{Convergence in terms of the truncation distance} \label{sec:sim:truncation}

Next, we investigate the convergence with respect to the real-space truncation distance $d$.
Here, the calculations are done at $\mu=-3.7$ and $-1.3$ with $m_{\rm tot}=40$, for which PEMC results show good convergence to the EDMC ones as discussed in the previous sections. 
The system size is $N=4\times4^3$, in which the Manhattan distance to the furthest site is $d=8$.

\begin{figure}
\begin{center}
\includegraphics[width=0.6\linewidth]{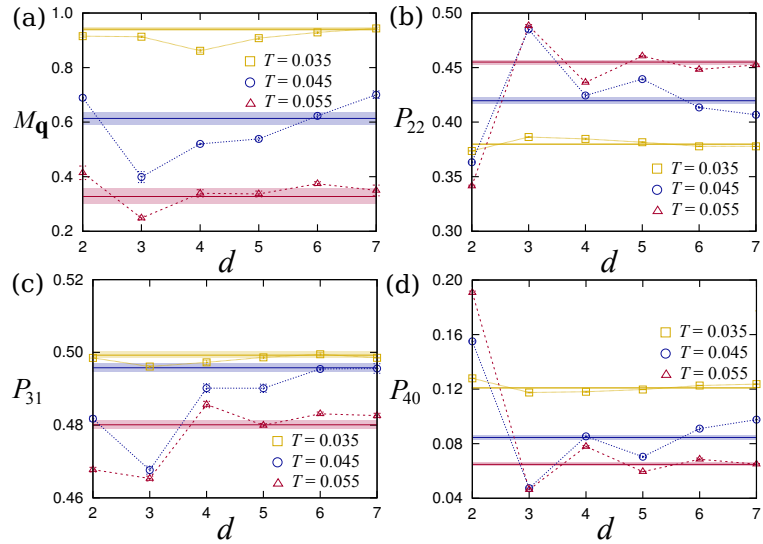}
\end{center}
\caption{
(color online). 
PEMC results for $d$ dependences of (a) $M_{\bf q}$, (b) $P_{22}$, (c) $P_{31}$, and (d) $P_{40}$. 
The calculations are done for $N=4\times 4^3$ and $\mu=-3.7$ with $m_{\rm tot}=40$.
For comparison, the results and statistical errors by EDMC are shown by horizontal solid lines and shades, respectively.
}
\label{fig:sim:npoly37d}
\end{figure}

Figure~\ref{fig:sim:npoly37d} shows the PEMC results in the 32-sublattice ordered region at $\mu=-3.7$ for different temperatures, $T=0.035$, $0.045$, and $0.055$, which correspond to the magnetically ordered, critical, and paramagnetic regions, respectively.
In all $T$ regions, the PEMC data converge to the EDMC ones for $d \gtrsim 6$, except for the data at $T=0.045$ in the critical region. 
This shows again that the convergence becomes slower in the critical region.

\begin{figure}
\begin{center}
\includegraphics[width=0.6\linewidth]{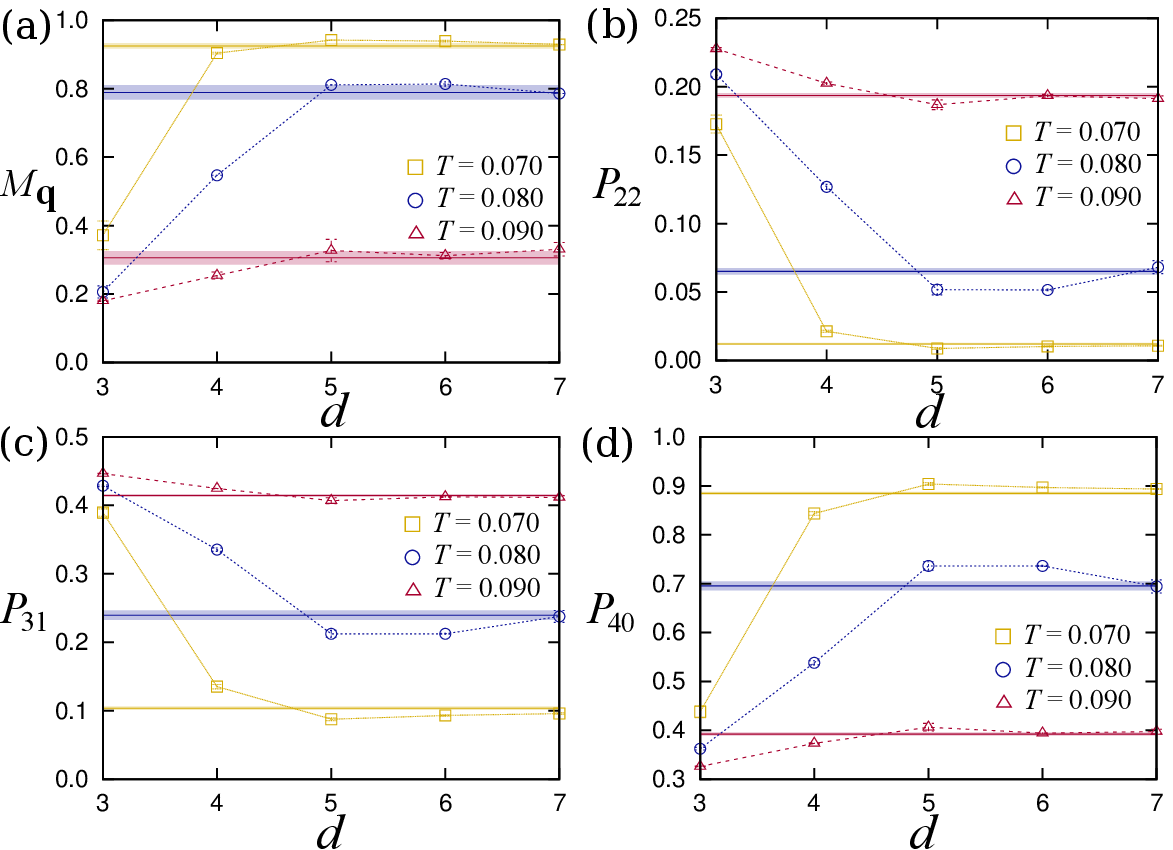}
\end{center}
\caption{
(color online). 
PEMC results for $d$ dependences of (a) $M_{\bf q}$, (b) $P_{22}$, (c) $P_{31}$, and (d) $P_{40}$. 
The calculations are done for $N=4\times 4^3$ and $\mu=-1.3$ with $m_{\rm tot}=40$.
For comparison, the results and statistical errors by EDMC are shown by horizontal solid lines and shades, respectively.
}
\label{fig:sim:npoly13d}
\end{figure}

The situation is similar in the all-in/all-out region.
Figure~\ref{fig:sim:npoly13d} shows the results for $\mu=-1.3$ at $T=0.070$, $0.080$, and $0.090$.
The results for $M_{\bf q}$ in Fig.~\ref{fig:sim:npoly13d}(a) show good convergence for $d \gtrsim 5$ for all $T$ shown.
For $P_{22}$, $P_{31}$, and $P_{40}$ in Figs.~\ref{fig:sim:npoly13d}(b)-\ref{fig:sim:npoly13d}(d), the results at $T=0.070$ and $0.090$ also show well converged results for $d \gtrsim 5$.
On the other hand, the data at $T=0.080$ shows a slight deviation from EDMC data up to $d \simeq 7$.

From the above results, it appears that the truncation with $d\gtrsim 6$ gives sufficient convergence in a wide range of $T$ and $n$.
In this system size $N=4\times 4^3$, $d=6$ already covers a large part of the lattice sites, and hence, the truncation is less useful to the reduction of the calculation amount. 
It is, however, expected from the truncation algorithm that the necessary truncation distance for the same accuracy does not depend so much on the system size; hence, the truncation will be efficient for larger system sizes.  
We examine this point in the next section.

\subsection{Convergence in terms of the truncation distance in larger systems} \label{sec:sim:largesize}

\begin{figure}
\begin{center}
\includegraphics[width=0.6\linewidth]{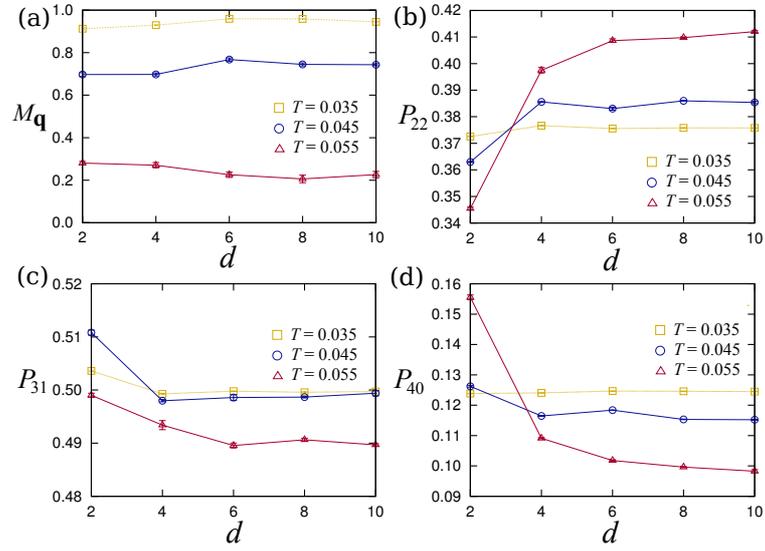}
\end{center}
\caption{
(color online). 
PEMC results for $d$ dependences of (a) $M_{\bf q}$, (b) $P_{22}$, (c) $P_{31}$, and (d) $P_{40}$. 
The calculations are done for $N=4\times 6^3$ and $\mu=-3.4$ with $m_{\rm tot}=40$.
}
\label{fig:sim:trunc6}
\end{figure}

To confirm the efficiency of the truncation in larger system sizes, here we conduct the PEMC calculations for $N=4\times 6^3$ and $4\times 8^3$ in the 32-sublattice ordered region.
For the system sizes, as EDMC is inapplicable due to the large calculation amount, we perform only PEMC and check the convergence with respect to the truncation distance $d$.
In these sizes, the Manhattan distance to the farthest site is $d=12$ for $N=4\times 6^3$ and $d=16$ for $N=4\times 8^3$.

\begin{figure}
\begin{center}
\includegraphics[width=0.6\linewidth]{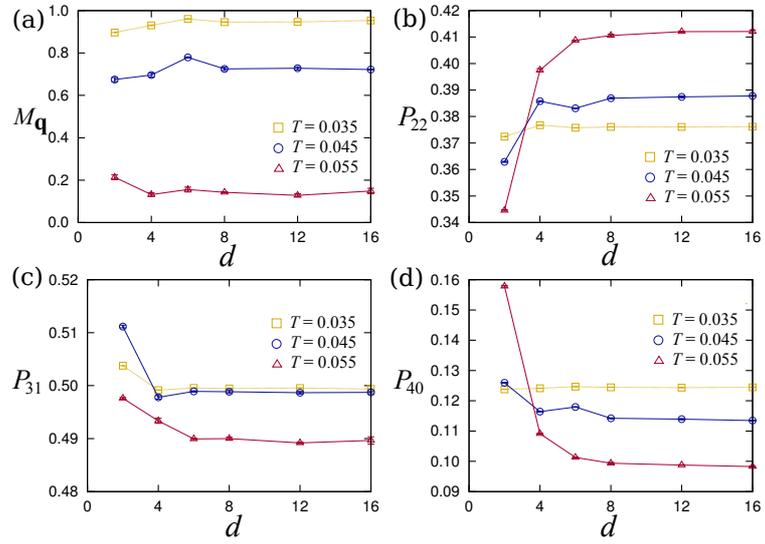}
\end{center}
\caption{
(color online). 
PEMC results for $d$ dependences of (a) $M_{\bf q}$, (b) $P_{22}$, (c) $P_{31}$, and (d) $P_{40}$. 
The calculations are done for $N=4\times 8^3$ and $\mu=-3.4$ with $m_{\rm tot}=40$.
}
\label{fig:sim:trunc8}
\end{figure}

Figure~\ref{fig:sim:trunc6} shows the results in the 32-sublattice ordered region at $\mu=-3.4$ [$n=0.195(1)$] for $N=4\times 6^3$.
Here, we take $m_{\rm tot}=40$, as the necessary $m_{\rm tot}$ for the convergence is expected to be less dependent on the system sizes~\citep{Motome1999,Furukawa2001}.
As shown in Fig.~\ref{fig:sim:trunc6}(a), for $d\gtrsim 8$, the results for $M_{\bf q}$ show reasonable convergence to the results without truncation in all $T$ regions.
The results for $P_{22}$, $P_{31}$, and $P_{40}$ also show convergence with $d\gtrsim8$, as shown in Figs.~\ref{fig:sim:trunc6}(b)-(d).
Similar behavior is also observed for $N=4\times 8^3$.
Figure~\ref{fig:sim:trunc8} shows the results for $N=4\times 8^3$ with $T=0.035$, $0.045$, and $0.055$.
All the results also show good convergence for $d\gtrsim8$. 

\section{Discussion} \label{sec:discussion}

Let us discuss our results in comparison with the previous ones on similar Kondo lattice models.
In the previous studies using PEMC, the number of polynomials for well converged calculations was typically $30\lesssim m_{\rm tot} \lesssim 40$~\citep{Motome1999,Alvarez2005,Sen2006,Zhang2011,Zhang2012}.
Our results presented at $\mu=-3.7$ and $-1.3$ in Sec.~\ref{sec:sim:Tdep} and Sec.~\ref{sec:sim:polynomial} indicate that good convergence is reached for similar range of $m_{\rm tot}$.
This shows that PEMC is also an efficient approach even in the presence of severe geometrical frustration. 
We note that the range of chemical potential $\mu$ corresponds to a moderate electron density $n \gtrsim 0.15$. 
Considering the fact that most of the previous studies were conducted in the region for $0.20\lesssim n\lesssim 0.80$~\citep{Motome1999,Alvarez2005,Sen2006,Zhang2011,Zhang2012}, this also supports the applicability of PEMC in the frustrated models in the similar density region. 

On the other hand, our results in the lower electron density region show much slower convergence, and even the results for $m_{\rm tot}=80$ show considerable deviations from the EDMC results, as shown in Fig.~\ref{fig:sim:npoly57m}.
This might be owing to the fact that the Fermi level is close to the band bottom.
In the small electron density region, the precise structure of DOS near the band edge plays a crucial role for the thermodynamics; to reproduce the details of DOS requires larger number of polynomials.
Another possible source is the small energy scale in the low density region. 
Because of the small kinetic energy, the effective interactions between localized spins become small, and hence, the relevant $T$ range including $T_c$ is much lower than that in the higher density region. 
This also requires larger number of polynomials for sufficient convergence.

Next, we discuss the convergence with respect to the real-space truncation.
The results in Figs.~\ref{fig:sim:trunc6} and \ref{fig:sim:trunc8} show that sufficient convergence is obtianed for $d \gtrsim 8$ for both $N=4\times 6^3$ and $4\times 8^3$.
This is consistent with the expectation that the necessary truncation distance is not strongly dependent on the system size. 
Unfortunately, as $d=8$ covers the large part of the system with $N=4\times 6^3$ and $4\times 8^3$, the truncation method is not helpful to reducing the calculation amount for the present system sizes.
It is, however, expected to be efficient for much larger system sizes if accessible.

\section{Summary} \label{sec:summary}

To summarize, we have presented the benchmark results on the application of polynomial expansion Monte Carlo method to a geometrically-frustrated spin-charge coupled system, a spin-ice type Kondo lattice model on a pyrochlore lattice.
We have investigated the convergence of Monte Carlo results with respect to the number of polynomials $m_{\rm tot}$ and the truncation Manhattan distance $d$. 
The results indicate that, in the electron density region $0.15 \lesssim n \lesssim 0.35$, the polynomial expansion Monte Carlo results with $m_{\rm tot}=40$ show sufficient convergence to those obtained by the conventional Monte Carlo method using the exact diagonalization. 
The results show that, although the current model has a $\delta$-function singularity in the density of states in the noninteracting limit associated with the geometrical frustration, the polynmial expansion Monte Carlo results show good convergence within the number of polynomials comparable to previous studies for unfrustrated models.
For the real-space truncation, our results indicate that $d\gtrsim 8$ gives well converged results for $N=4\times 6^3$ and $4\times 8^3$, while $d\gtrsim 6$ is enough for $N=4\times 4^3$.
Although the truncation algorithm is not useful for reducing the calculation amount for the system sizes that we calculated, the small system-size dependence of the necessary truncation distance implies that the truncation becomes efficient for larger system sizes.

Recent experiments in the metallic pyrochlore oxides have stimulated the studies on spin-charge coupled systems on frustrated lattices.
There, the polynomial expansion Monte Carlo method will be a powerful theoretical tool if it is not suffered from the difficulty specific to the frustrated systems, the pathological singularity in the density of states.
Our results indicate that, in contrast to what was anticipated, the polynomial expansion Monte Carlo method is efficiently applied to the frustrated systems.
As the method provides numerically exact solutions, the present benchmark will further stimulate numerical studies on the spin-charge coupled systems on frustrated lattices.

\section{acknowledgment}

The authors thank T. Misawa and H. Shinaoka for fruitful discussions and helpful comments.
Part of the calculations were performed on the Supercomputer Center, Institute for Solid State Physics, University of Tokyo.  H.I. is supported by Grant-in-Aid for JSPS Fellows.
This research was supported by KAKENHI (No.19052008, 21340090, 21740242, 22540372, 24340076, and 24740221), Global COE Program ``the Physical Sciences Frontier", the Strategic Programs for Innovative Research (SPIRE), MEXT, and the Computational Materials Science Initiative (CMSI), Japan.

\bibliographystyle{elsarticle-harv}
\bibliography{<your-bib-database>}







\end{document}